\documentclass[prb,twocolumn,showpacs]{revtex4}

\usepackage{graphics}
\usepackage{graphicx}
\usepackage{amsmath}
\usepackage{amstext}
\usepackage{amssymb}
\usepackage{hyperref}
\def  \bsig   {\mbox{\boldmath$\sigma$}}

\def \bmatA   {\mbox{\boldmath$\mathcal{A}$}}

\begin{document}

\title{Chiral two-dimensional electron gas in a periodic magnetic field}

\author{M. Taillefumier$^{1,2,3}$, V. K. Dugaev$^{4,5}$, B. Canals$^2$,
C. Lacroix$^2$, and P. Bruno$^1$}

\affiliation{$^1$Max-Planck-Institut f\"ur Mikrostrukturphysik,
Weinberg 2, 06120 Halle, Germany\\
$^2$Institut N\'eel, CNRS/UJF, 25 avenue des Martyrs, BP 166,
38042 Grenoble, Cedex 09, France\\
$^3$Department of Physics, Norwegian University of Science
and Technology, N-7491 Trondheim, Norway\\
$^4$Department of Physics, Rzesz\'ow University of Technology,
al. Powsta\'nc\'ow Warszawy 6, 35-959 Rzesz\'ow, Poland\\
$^5$Department of Physics and CFIF, Instituto Superior T\'ecnico,
Universidade T\'ecnica de Lisboa, Av. Rovisco Pais, 1049-001 Lisbon, Portugal}

\date{\today }

\begin{abstract}
We study the energy spectrum and electronic properties of
two-dimensional electron gas in a periodic magnetic field of zero
average with a symmetry of triangular lattice.
We demonstrate how the structure of electron energy bands can be
changed with the variation of the field strength, so that we can
start from nearly free electron gas and then transform it
continuously to a system of essentially localized chiral electron
states.
We find that the electrons near some minima of the
effective potential are responsible for occurrence of dissipationless
persistent currents creating a lattice of current
contours.
The topological properties of the electron energy bands
are also varied with the intensity of periodic field.
We
calculated the topological Chern numbers of several lower energy
bands as a function of the field.
The corresponding Hall conductivity is nonzero and, when the Fermi
level lies in the gap, it is quantized.
\end{abstract}
\pacs{73.21.-b,73.50.Jt,75.47.-m,73.23.Ra}
\maketitle

\section{Introduction}

The aim of this work is to study the effect of a periodic magnetic
field of zero average on the dynamic of a free electron gas.
The magnetic field distribution forms a "magnetic-field lattice" for
electrons, which results in the formation of electron energy bands
controlled by the strength and the geometry of the magnetic field.

The possibility of using periodic magnetic fields for tailoring the
electronic structure is mostly related to recent advances in
nanotechnology, which enables to manufacture two-dimensional lattices
of ferromagnetic nanocylinders\cite{Nielsch2001}.
This
  idea has been already used by us to suggest a system where the spin
  chirality mechanism related to the anomalous Hall effect (AHE) in
  frustrated
  ferromagnets\cite{Ye1999,Taguchi2001,Tatara2002,S_Onoda2003} can be
  measured and controlled externally~\cite{Bruno2004}. To detect this
  effect, we proposed to measure the Hall effect in 2D diluted
  magnetic semiconductor on top of the nanolattice of ferromagnetic
  cylinders.
Another possible way to create the periodic field is to use an array
of magnetic nanodots with tunable out-of-plane magnetization like in
Ref.~[\onlinecite{Silhanek2007}].

Previous investigations of the 2D electronic system in periodic
magnetic fields concentrated mostly on the one-dimensional periodic
modulation\cite{Ibrahim1995,Gu1997} and, in some cases, on a mutual
effect of the uniform and periodic magnetic
fields.\cite{Xue1992,Krakovsky1996,Shi1997,Chang1994,Rom1996} The main
difference of our work is that we assume that the uniform magnetic
field is exactly zero, whereas the periodic field forms a real
two-dimensional nanolattice. For definiteness, here we focus on the
case of triangular lattice, which corresponds to the nanocylinder
structure of Ref.~[\onlinecite{Nielsch2001}]. It should be emphasized
that the assumption of zero uniform field is very important because it
results in formation of well-defined electron energy bands
characterized by the electron momentum ${\bf k}$ like in the case of
electric modulation. On the other hand, the presence of 2D
magnetic-field modulation substantially changes a picture of the
"snakelike" electron motion in a nonuniform (linear-in-gradient)
magnetic field.\cite{Muller1992}

The semiclassical consideration of the motion of electrons in
inhomogeneous magnetic field shows that the low-energy electrons
mostly tend to localize near the lines of zero magnetic
field.\cite{Muller1992,Hofstetter1996} 
The corresponding effective potential has a different form for
electrons moving in opposite directions along the zero-field line.
We found an analogous tendency to localization of electrons in the
2D periodic field. In this case, the zero-field lines correspond
to some closed trajectories of electron motion, which is chiral
and quantized and which leads to occurrence of equilibrium {\it
persistent currents}. It should be noted that usually the
persistent currents are associated with mesoscopic rings, for
which the symmetry of the electron motion in opposite directions
is broken by the magnetic
field.\cite{Buttiker1983,Cheung1988,Altshuler1991,Loss1991,Fujimoto1993,Ding2007}
However, in our case of the {\it "magnetic crystal"}, the persistent
currents appear like a periodic array of circular currents.

Quite recently a lot of discussions has been induced by the study of
an "intrinsic" mechanism of the anomalous Hall effect (AHE).
\cite{Jungwirth2002,M_Onoda2002} In the ballistic regime, when the
impurities can be totally neglected, the anomalous Hall effect is
related to the topology of electron energy bands, which can be
characterized by integers called Chern numbers. It turned out that the
discussion of the intrinsic mechanism of AHE lead to a partial
revising of the Landau theory of Fermi liquid because the transport
properties are found to be related to the Berry curvature of electron
bands in momentum space,\cite{Sundaram1999} which means that the
corresponding topological element should be added to the Landau
theory.\cite{Haldane2004}

Some rather simplified theoretical models like the 2D electron gas
with Rashba spin-orbit interaction or the relativistic 2D Dirac model
allow full analytical calculation of the Berry curvatures and Chern
numbers.
Recently, several publications reported
  calculations of the Berry curvature for SrRuO$_3$\cite{Fang2003} and
  bcc Fe\cite{Yao2004,Wang2006} using {\it ab initio} methods.
In our previous publication,\cite{Taillefumier2006} we presented numerical
calculations of the Berry curvature of the energy bands of electrons
interacting with a chiral spin texture defined on top of a kagom\'e
lattice.
However, in the latter case the essential element is the
inhomogeneous orientation of the localized spins, which leads to the
chirality contribution\cite{Ye1999,Taguchi2001,Tatara2002,S_Onoda2003}
to the AHE.

In this work, we study the topology of electron energy bands by
calculating the Chern numbers as a function of the intensity of
periodic magnetic field. The model with the periodic field gives us
such a parameter, which can presumably be varied in the experiment
(for example, by changing a distance between the 2D layer and the
nanocylinder lattice). This way we can demonstrate a strong jump-like
dependence of the Chern numbers on the field.

\section{Model}

We consider a model of two-dimensional electron gas (2DEG)
of spinless electrons in the $x-y$ plane under periodic magnetic
field ${\bf B}({\bf r})$. Since the electrons move in the plane,
the only component of field acting on electrons
is the component perpendicular to this plane, which we denote by
$B({\bf r})$. Then the properties of the system can be
described by the following Hamiltonian:
\begin{equation}
  \label{eq:1}
  \mathcal{H}=\frac{\hbar ^2}{2 m}\left(-i{\bf\nabla}
    -\frac{e}{\hbar c} {\bf A}({\bf r})\right)^2,
\end{equation}
where $m$ is the effective electron mass and ${\bf A}({\bf r})$ is
the vector potential related to the magnetic field $B({\bf r})$ by
$\nabla \times A({\bf r})=B({\bf r})$. In the following we use the
Coulomb gauge determined by the condition ${\bf\nabla}\cdot {\bf
A}({\bf r})=0$. In the case of zero average field, it is possible
to choose the vector
potential ${\bf A}({\bf r})$ as periodic-in-space (see below).

Thus, we have a problem of electron in a periodic potential, and
we can use the Bloch theorem for the eigenfunctions of Hamiltonian
(\ref{eq:1})
\begin{equation}
  \label{eq:2}
  \psi_{n,\bf k}({\bf r})=e^{i {\bf k}\cdot {\bf r}} u_{n,\bf k}({\bf r}),
\end{equation}
where $n$ is the band index and ${\bf k}$ is the crystal momentum,
which is restricted to the first Brillouin zone. The function
$u_{n,\bf k}({\bf r})$ is periodic, $u_{n,\bf k}({\bf r})=u_{n,\bf
k}({\bf r}+{\bf R})$ with ${\bf R}$ being the lattice vector
determined by the periodicity of the potential. Substituting
Eq.~(\ref{eq:2}) in Eq.~(\ref{eq:1}), we find that $u_{n,\bf
k}({\bf r})$ verify
\begin{equation}
  \label{eq:3}
  \mathcal{H}_{\bf k}\, u_{n,\bf k}({\bf r})
  =\varepsilon_{n,\bf k}\, u_{n,\bf k}({\bf r}),
\end{equation}
where $\varepsilon_{n,\bf k}$ is the eigenvalue associated with
the eigenfunction $\psi_{n,\bf k}({\bf r})$ and
\begin{equation}
  \label{eq:4}
  \mathcal{H}_{\bf k}=\frac{\hbar ^2}{2
    m}\left(-i{\bf\nabla}+{\bf k}
    -\frac{e}{\hbar c} {\bf A}({\bf r})\right)^2
\end{equation}
is the reduced Hamiltonian depending on ${\bf k}$. Using the
periodicity of $u_{n,\bf k}({\bf r})$, we can write it as
\begin{equation}
  \label{eq:5}
  u_{n,\bf k}({\bf r})=\sum_{\bf g} u_{n,\bf k} ({\bf g})\, e^{i {\bf
      g}\cdot {\bf r}},
\end{equation}
where ${\bf g}$ is a vector of the reciprocal lattice, $u_{n,\bf
  k}({\bf g})$ is the Fourier transform of $u_{n,\bf k}({\bf r})$
defined by
\begin{equation}
  \label{eq:6}
  u_{n,\bf k}({\bf g})=\frac{1}{S}\int_{s} u_{n,\bf k}({\bf r})\,
  e^{-i {\bf g}\cdot {\bf r}} d^2{\bf r},
\end{equation}
and $S$ is the area of the unit cell. Substituting
Eq.~(\ref{eq:6}) into (\ref{eq:3}), we find that $u_{n,\bf
  k}({\bf g})$ satisfy the following set of equations
\begin{equation}
  \label{eq:7}
  \sum_{\bf g^\prime}\mathcal{H}_{\bf k}({\bf g},{\bf g}^\prime)\, u_{n,\bf k}({\bf
    g^\prime}) = \varepsilon_{n,\bf k} u_{n,\bf k}({\bf g}).
\end{equation}
The matrix elements $\mathcal{H}_{\bf k}({\bf g},{\bf g}^\prime)$
can be calculated by using Eqs.~(\ref{eq:4}) and (\ref{eq:5}):
\begin{eqnarray}
  \label{eq:8}
  \mathcal{H}_{\bf k}({\bf g},{\bf
    g}^\prime)&=&\frac{\hbar^2}{2m}\left[({\bf k}+{\bf
      g})^2\delta_{{\bf g},{\bf g}^\prime}
      -\frac{2 e}{\hbar c}({\bf k}+{\bf g})\cdot {\bf A}({\bf
      g}-{\bf g}^\prime)\right.\nonumber\\
  &+&\left.\frac{e^2}{\hbar^2 c^2}\; {\bf A}^2({\bf g}-{\bf g}^\prime)\right] ,
\end{eqnarray}
where ${\bf A}({\bf g})$ and ${\bf A}^2({\bf g})$ are the Fourier
transforms of ${\bf A}({\bf r})$ and ${\bf A}^2({\bf r})$,
respectively. Note that Eq.~(\ref{eq:8})
depends on the gauge, and is written in the
Coulomb gauge ${\bf g}\cdot {\bf A}({\bf g})=0$.
Finally, the Fourier transform of ${\bf
  A}^2({\bf r})$ is related to ${\bf A}({\bf g})$ by the convolution
\begin{equation}
  \label{eq:9}
  {\bf A}^2({\bf g})=\sum_{{\bf g}^\prime} {\bf A}({\bf g}-{\bf
    g}^\prime)\cdot {\bf A}({\bf g}).
\end{equation}

Up to now, the derivation is quite general. In the following, we
specify the form of the magnetic field profile to the symmetry of
triangular lattice. It can be realized by using a periodic array
of ferromagnetic nanocylinders with a period of few ten
nanometers\cite{Nielsch2001} on top of 2DEG, like explained in
Ref.~[\onlinecite{Bruno2004}]. This distribution is periodic in
plane and its Fourier components are decreasing exponentially with
the wave vector ${\bf g}$ and with the distance between the 2D
film and the array of ferromagnetic nanocylinders.

\begin{figure}[ht]
\centering
\includegraphics[width=7cm]{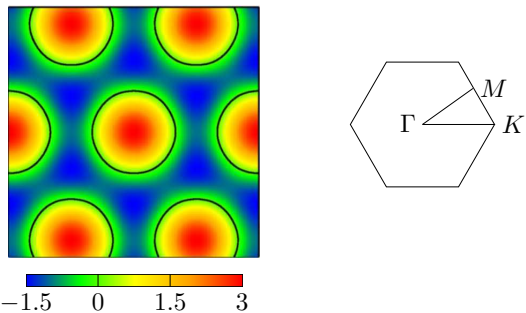}
\caption{(Color online) (Left panel) Density plot of the magnetic
  field distribution (in arbitrary units). The black lines are the
  isolines $B(\bf r)=0$. (Right panel) First Brillouin zone of the
  triangular lattice.}
\label{Fig:1}
\end{figure}

The magnetic field has a zero net flux over the unit cell and we
assume that only the first Fourier components of the field are
important. The magnetic field profile is shown in
Fig.~\ref{Fig:1}. Our approximation is valid when the distance
between the 2DEG and the array of ferromagnetic nanocylinders is
of the order of the lattice period.

We write down the $z$-component of the magnetic field as
\begin{eqnarray}
  \label{eq:10}
  B({\bf r})=B_0 \left[\cos\left(\frac{2\pi}{a}\, {\bf b}_1 \cdot {\bf
        r}\right)
        +\cos\left(\frac{2\pi}{a}\, {\bf b}_2 \cdot {\bf
        r}\right)\right.\nonumber\\
  +\left.\cos\left(\frac{2\pi}{a}\, {\bf b}_3 \cdot {\bf
        r}\right)\right],
\end{eqnarray}
where $B_0$ is the amplitude of the field and $a$ is the lattice
period. The vectors ${\bf b}_i$ with $i=1,2$ and ${\bf b}_3={\bf
  b}_2+{\bf b}_1$ are the vectors of reciprocal lattice of the
triangular lattice. They are defined by the relations
${\bf b}_i\cdot {\bf a}_j=\delta_{ij}$, where $i,j=1,2$, ${\bf
  a}_1=(1,0)$, ${\bf a}_2=(1/2,\sqrt{3}/2)$, and ${\bf a}_3={\bf
  a_2}-{\bf a}_1$.
  The first
Brillouin zone (Fig.~\ref{Fig:1}) is an 
hexagon with corners at ${\bf k}_i=\frac{4\pi}{3 a} {\bf a}_i$.
The (Coulomb gauge) vector potential ${\bf A}({\bf r})$ can be
chosen in the following form (i.e., periodic in space)
\begin{eqnarray}
  \label{eq:11}
  {\bf A}({\bf r})=A_0\left[{\bf a_2}\sin\left(\frac{2\pi}{a}\, {\bf
        b}_1\cdot {\bf r}\right)
        -{\bf a}_1\sin\left(\frac{2\pi}{a}\, {\bf
        b}_2\cdot {\bf r}\right)\right.\nonumber\\
  +\left.{\bf a}_3 \sin\left(\frac{2\pi}{a}\, {\bf
        b}_3\cdot {\bf r}\right)\right],\hskip0.3cm
\end{eqnarray}
where $A_0= B_0\sqrt{3} a/4\pi$.

\begin{figure}[ht]
  \includegraphics[width=6.4cm]{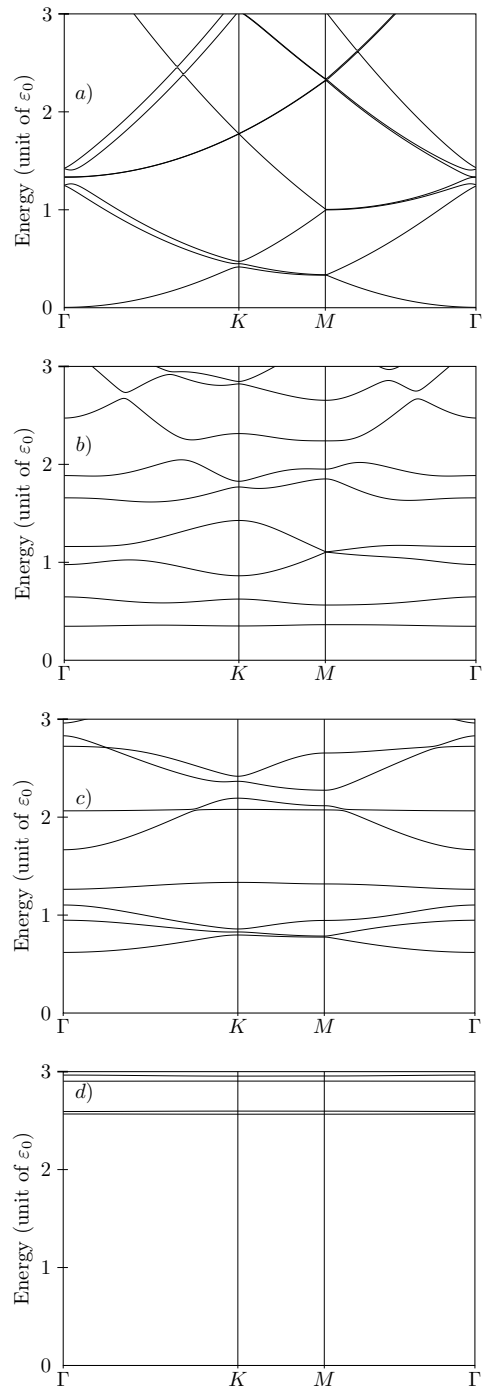}
  \caption{Energy spectrum of electrons in periodic magnetic field
    calculated for $\alpha=0.05$ (a), $0.5$ (b), $1$ (c), and $5$
    (d). Local gaps are open at points of the Brillouin zone where
    degeneracies due to the Bragg plan are located.}
  \label{fig:2}
\end{figure}

For the numerical calculations we define the reduced units $B({\bf
  r})\to B_0 B({\bf r})$, ${\bf A}({\bf r})\to A_0 {\bf A}({\bf r})$,
${\bf g}\to \frac{2 \pi}{a} {\bf g}$ is the vector of reciprocal
lattice and ${\bf k}\to \frac{2\pi}{a} {\bf k}$ the crystal
momentum taken in the first Brillouin zone. The energy
$\varepsilon$ is replaced by $\varepsilon\to \varepsilon_0
\varepsilon$ with $\varepsilon_0=\hbar^2/2 ma^2$. Finally, after
introducing the dimensionless parameter $\alpha= -e A_0 a/ hc$,
Eq.~(\ref{eq:8}) becomes
\begin{eqnarray}
  \label{eq:12}
  \mathcal{H}_{\bf k}({\bf g},{\bf g}^\prime)
  =\varepsilon_0\left[({\bf k}+{\bf g})^2\delta_{{\bf g},{\bf g}^\prime}
  +2\alpha{\bf A}({\bf g}-{\bf g}^\prime)\cdot ({\bf k}+{\bf g})
  \right.\nonumber\\
  +\left.\alpha^2 {\bf A}^2({\bf g}-{\bf g}^\prime)\right] ,\hskip0.5cm
\end{eqnarray}
so that unless stated, all the quantities will be presented in the
reduced dimensionless units. In what follows $\alpha $ is the main
parameter determining the intensity of the magnetic-field modulation.

\section{Electron energy spectrum and wave functions}

\subsection{Weak magnetic field: perturbation theory}\label{Sec:energy:weak}

Due to our choice of the gauge for the vector potential (\ref{eq:11}),
one can use the perturbation theory over ${\bf A}({\bf r})$, which
corresponds to the limit of weak magnetic field. Then the unperturbed
Hamiltonian is $\mathcal{H}_0=-\hbar ^2\nabla ^2/2m$ (in this section
we do not use reduced units), and the Hamiltonian of interaction
\begin{equation}
\label{13}
\mathcal{H}_{int}=\frac{ie\hbar }{mc}\, {\bf A}({\bf r})\cdot \nabla
+\frac{e^2}{2mc^2}\, {\bf A}^2({\bf r})
\end{equation}
with
\begin{equation}
\label{14}
{\bf A}({\bf r })={\bf A}_2\sin ({\bf g}_1\cdot {\bf r})
-{\bf A}_1\sin ({\bf g}_2\cdot {\bf r})
+{\bf A}_3\sin ({\bf g}_3\cdot {\bf r}),
\end{equation}
where ${\bf A}_i=A_0{\bf a}_i$ and ${\bf g}_i=\frac{2\pi }{a}{\bf
  b}_i$. The eigenfunctions of $\mathcal{H}_0$ are the usual plane
waves $\psi _{\bf k}({\bf r})=\frac1{2\pi }e^{i{\bf k}\cdot {\bf r}}$,
and the corresponding matrix elements of interaction can be calculated
as the Fourier components of $\mathcal{H}_{\text{int}}$
\begin{eqnarray}
\label{15}
\mathcal{H}_{\text{int}}({\bf q})=\frac{e^2A_0^2}{8mc^2}\left[
6\, \delta ({\bf q})-\delta ({\bf q}+2{\bf g}_1)
-\delta ({\bf q}-2{\bf g}_1)\hskip0.5cm
\right. \nonumber \\ \left.
-\delta ({\bf q}+2{\bf g}_2)
-\delta ({\bf q}-2{\bf g}_2)
-\delta ({\bf q}+2{\bf g}_3)
-\delta ({\bf q}-2{\bf g}_3)
\right. \nonumber \\ \left.
+\delta ({\bf q}+{\bf g}_1+{\bf g}_2)
-\delta ({\bf q}+{\bf g}_1-{\bf g}_2)
-\delta ({\bf q}-{\bf g}_1+{\bf g}_2)
\right. \nonumber \\ \left.
+\delta ({\bf q}-{\bf g}_1-{\bf g}_2)
-\delta ({\bf q}+{\bf g}_1+{\bf g}_3)
+\delta ({\bf q}+{\bf g}_1-{\bf g}_3)
\right. \nonumber \\ \left.
+\delta ({\bf q}-{\bf g}_1+{\bf g}_3)
-\delta ({\bf q}-{\bf g}_1-{\bf g}_3)
-\delta ({\bf q}+{\bf g}_2+{\bf g}_3)
\right. \nonumber \\ \left.
+\delta ({\bf q}+{\bf g}_2-{\bf g}_3)
+\delta ({\bf q}-{\bf g}_2+{\bf g}_3)
-\delta ({\bf q}-{\bf g}_2-{\bf g}_3) \right] .\hskip0.3cm
\end{eqnarray}
It should be noted that any matrix elements of the first (linear in
${\bf A}$) term in Eq.~(\ref{13}) are zero.

The perturbation related to $\mathcal{H}_{\text{int}}$ breaks the
degeneracy of states belonging to the points in the Brillouin zone
separated by vector ${\bf g}$, for which $\mathcal{H}_{\text{int}}({\bf
g})\ne 0$. Using Eq.~(\ref{15}) we find the matrix elements
corresponding to transitions between the states in the opposite
points at the Brillouin zone edges
\begin{eqnarray}
\label{16}
\mathcal{H}_{\text{int}}({\bf g}_1)=\mathcal{H}_{\text{int}}({\bf g}_2)
=\mathcal{H}_{\text{int}}({\bf g}_3)=\frac{e^2A_0^2}{8mc^2}\; .
\end{eqnarray}
For example, the element $\mathcal{H}_{\text{int}}({\bf g}_2)$ couples
the degenerate states ${\bf k}=(0,\, -2\pi /a\sqrt{3})$ and ${\bf
k}^\prime =(0,\, 2\pi /a\sqrt{3})$. Using the perturbation theory
for the degenerate states ${\bf k}$ and ${\bf k}^\prime $, we can
find that the value $e^2A_0^2/8mc^2$ determines the magnitude of
the corresponding energy gap at the Brillouin zone edge. Note that
the gap in these points is nonzero for any weak perturbation and it increases with
the amplitude of magnetic field as $B_0^2$.

We can find that the perturbation theory approach is valid for
$|\alpha |\ll 1$. This condition can be also presented as $\phi /\phi
_0\ll 1$, where $\phi =B_0a^2\sqrt{3}/2$ is the flux
of field $B_0$ per elementary cell and $\phi _0=hc/e$ is the flux
quantum.

\subsection{Energy Spectrum}

\subsubsection{Degeneracies and symmetries of the hamiltonian}\label{sec:group:theory}

The Hamiltonian~(\ref{eq:1}) is invariant under discrete
translations of vectors ${\bf R}=i {\bf a}_1+j {\bf a}_2$ and
because of the polar nature of the vector potential ${\bf A}({\bf
r})$, it is also invariant under the point group $C_6$ (but not
$C_{6v}$) of pure six fold rotations.
Its space group is therefore abelian and its irreducible representations
are all of dimensions 1.
This physically means that the energy spectrum can only have accidental
degeneracies between consecutive energy bands.

\subsubsection{Energy spectrum: Numerical results}\label{Sec:energy:spectrum}

The solution of Eq.~(\ref{eq:1}) should be obtained by
diagonalizing the infinite matrix with elements given by
Eq.~(\ref{eq:8}). In practice, one cuts the basis to get a finite
matrix. We cut the basis by introducing the energy cutoff
$\varepsilon_c$, i.e., we keep the plane waves with energies
$\varepsilon\leq\varepsilon_c$ and neglect the others. Then we
diagonalize the obtained matrix by using the Lanczos algorithm
(with reorthogonalisation) implemented in the library
SLEPC\cite{slepc}.
 Finally, the
value of $\varepsilon_c$ is chosen to get the converged
quantities.

The presence of linear term, which couples the momentum ${\bf k}$ to
the vector potential, gives rise to a rich energy spectrum when the
amplitude of magnetic field is changed. In the low-field regime the
band structure (Fig.~\ref{fig:2},a) is similar to the band structure
of free particle. It is slightly modified near the points obeying
relation ${\bf k}^2=({\bf k}+{\bf g})^2$ because the Bloch band are
degenerated at these points. As we found in Sec.~\ref{Sec:energy:weak}
and Sec.~\ref{sec:group:theory}, the application of
a periodic vector potential leaves
the degeneracy.

The energy spectrum demonstrates a strong variation when the amplitude
of magnetic field is increased. The gaps can be seen for rather weak
values of $|\alpha |$ (Fig.~\ref{fig:2},b) and the band width
continuously decreases when $|\alpha |$ increases. Figures
\ref{fig:2},~b and c show that the band crossing occur at some high
symmetry points of the first Brillouin zone like the points $K,\; M$
and $\Gamma$. We see that the band crossing occur very often when
$|\alpha |$ is increased. As we shall see later, this property is
important for the quantization of Hall conductivity because it gives
rise to the jumps in $\sigma_{xy}$. The corresponding jumps of
$\sigma_{xy}$ when an external parameter is changed can be identified
as {\it topological transitions}.

Finally, for larger values of $|\alpha |$ (Fig.~\ref{fig:2}-d),
the bands are practically flat indicating that (i) electrons are
mostly localized and (ii) they have very low group velocity (i.e.,
the electrons are extremely heavy).

\subsection{Properties of the Bloch states}\label{sec:prop-bloch-stat}

Here we present the probability distribution of the Bloch states
in dependence on the parameter $\alpha$. Our study is restricted
to the high symmetry points of the first Brillouin zone, where the
Bloch state $u_{n,\bf k}({\bf r})$ has the symmetry of the point
${\bf k}$.

\begin{figure}[ht]
  \centering
  \includegraphics[width=7.5cm]{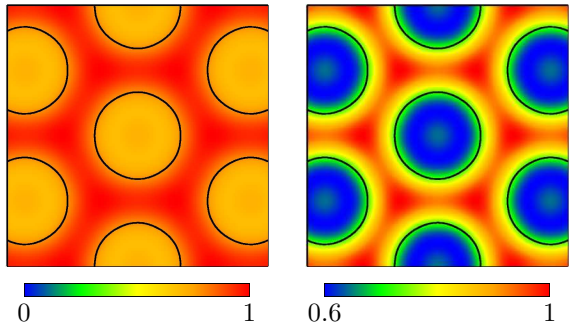}
  \caption{(Color online) Probability distribution for the Bloch
    states of the first energy band in $\Gamma$ point for
    $\alpha=0.5$. The black lines are the isolines of constant
    magnetic field $B(\bf r)=0$. The left panel shows that the
    particle is substantially delocalized over the unit cell, but
    avoids the regions where the magnetic field is maximum. This
    effect can clearly be seen on the right panel, where the color
    scale has been changed in order to see better the
    fine structure of the Bloch states.}
\label{fig:4}
\end{figure}

\begin{figure}[ht]
  \centering
  \includegraphics[width=80mm]{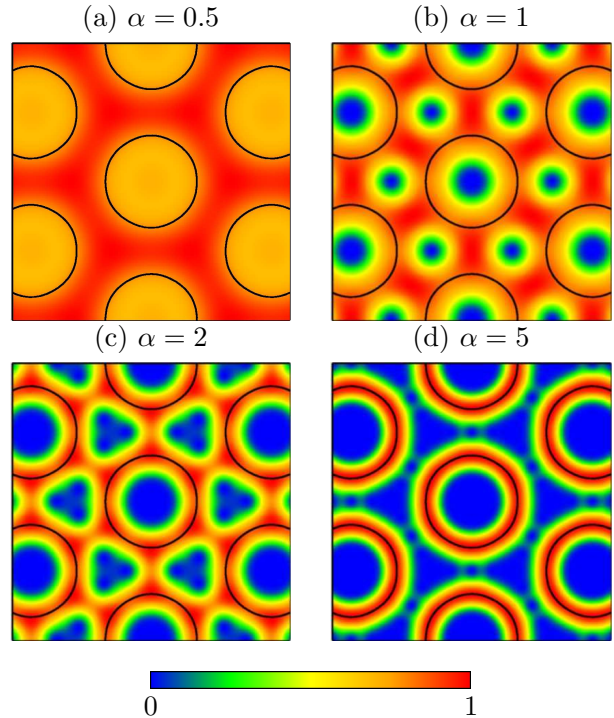}
  \caption{(Color online) Bloch states of the first band in the $\Gamma$
    point for (a) $\alpha=0.5$, (b) $\alpha=1$, (c) $\alpha=2$ and
    (d) $\alpha=5$. The black lines correspond to
    $B({\bf r})=0$. These figures shows that the particle tries to
    avoid the region where the magnetic field is large. This
    effect can be seen already in figure (a), but it is better
    visible in (b) and (c). For larger $|\alpha|$,
    the particle is confined around the lines of zero magnetic
    field.}
   \label{fig:5}
\end{figure}

Let us start by considering the $\Gamma$ point. The Bloch state at
this point has the symmetry of the lattice.
We consider first the band $n=1$ as a function of $\alpha$. The
results of calculation of the probability distribution $|\psi_{n,
\bf k}({\bf r})|^2$ for $\alpha =0.5$ are presented in
Fig.~\ref{fig:4}. At small $|\alpha |$ (see Fig.~\ref{fig:4}, left),
the electrons are mainly delocalized over the unit cell. However,
they avoid the regions where the magnetic field is large (inside
the regions delimited by the black lines on the figure). This is
also clearly illustrated by Fig.~\ref{fig:4}, right, where the color
scale has been changed in order to see the fine structure of the
distribution. As shown in Fig.~\ref{fig:5},~b, the increase of the
parameter $|\alpha |$ enhances this effect. The particles are
rejected from the region where the magnetic field is large and
concentrated in the regions where the field is close to zero
(Fig.~\ref{fig:5}, b,c). Figure~\ref{fig:5},~d represents the
limiting case where the repulsion effect confines the particle to
the region close to the line of $B({\bf r})=0$ (black lines in
Fig.~\ref{fig:5},~d). In this case, the particle is moving in an
effective potential created by the field profile, which
forms a ring, with the ring width depending on $\alpha$.

This behavior corresponds to the semiclassical picture of the
electron motion in linear magnetic field\cite{Muller1992}. In this
approach the low-energy electron drifts along the line of minimum
magnetic field, and the trajectory of this motion can be wavy or
snakelike depending on the drift direction. One can also
understand the effect of electron localization as a tendency to
occupy the region, in which the energy of Landau level is minimum.

\begin{figure}[ht]
  \centering
  \includegraphics[width=75mm]{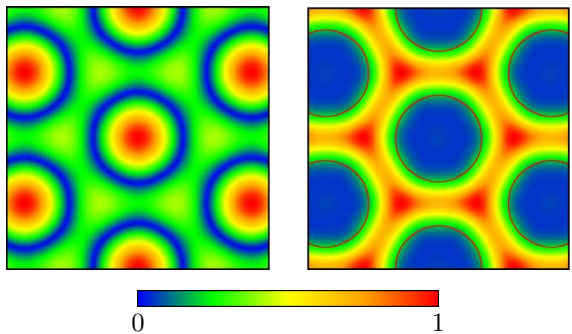}
  \caption{(Color online) Bloch states of the fifth band
    ($\alpha=0.5$) and fourth band ($\alpha=1$) in the $\Gamma$
    point. The particle is confined in the regions of strong magnetic
    field.}
   \label{fig:6}
\end{figure}

Up to now, we considered the states, in which the particle is
rejected into the regions of the weak field but one can also
obtain the states with particles rejected from these regions and
concentrated in the regions where $B(\bf r)$ is large. Such
situation can be found by considering higher energy bands. A
typical example is presented in the right panel of
Fig.~\ref{fig:6}.

We also calculated the probability distribution at other
symmetry points of the Brillouin Zone.
Except for a partial loss of symmetry, these electronic states
have properties similar to the electronic states at the $\Gamma$
point.

Our main result of this section is that the strong periodic field
results in localization of electrons in some low-energy states.
The electrons of the lowest energy bands are effectively confined
within some rings near the
closed lines of zero magnetic field, and the rings form a regular
array corresponding to the symmetry of the magnetic-field lattice.
The characteristic thickness of the rings decreases with the field
intensity.

\section{Persistent currents}\label{sec:exist-pers-curr}

Now we show that the electrons, which are confined within the
rings, are moving along the zero-field lines creating a regular
array of equilibrium persistent currents. For this purpose we
calculate the local current density ${\bf J}_{n,\bf k}(\bf r)$
defined as
\begin{equation}
  \label{eq:17}
  {\bf J}_{n,\bf k}({\bf r})=\frac{\hbar}{2 m}{\rm Re}\left[u_{n,\bf
      k}^\dagger({\bf r})\left(-i \nabla + {\bf k}
      -\frac{e}{\hbar} {\bf A}({\bf r})\right) u_{n, \bf k}({\bf r})\right] ,
\end{equation}
where $u_{n,\bf k}({\bf r})$ refers to the corresponding Bloch state.
One can also calculate the total current density ${\bf J}(\bf r)$
defined as the sum over all occupied states below the Fermi level
$\varepsilon_F$.

\begin{figure}
  \centering
   \includegraphics[width=75mm]{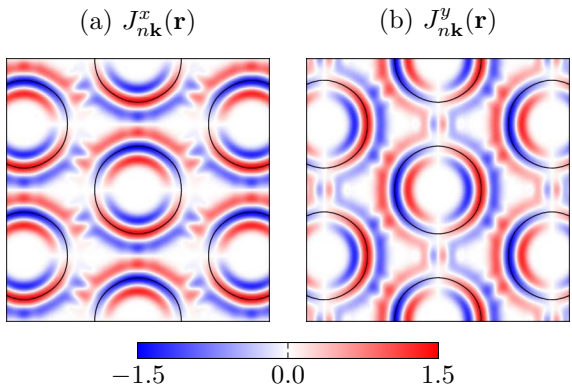}
   \caption{(Color online) Components of the current density (in unit
     of $h e/2 m a$) calculated for $\alpha = 5$ and $n=1$. It shows
     that the particle moves along the lines ${\bf B}({\bf r})=0$
     indicated in black.}
  \label{fig:9}
\end{figure}

Let us consider the current distribution at the $\Gamma$ point for
$\alpha=5$ and $n=1$. The distribution density for the corresponding
Bloch state is shown in Fig.~\ref{fig:5}-d, whereas Fig.~\ref{fig:9}
represents the spatial distribution of $x$ and $y$ components of
current density (\ref{eq:17}). As we see from Fig.~\ref{fig:9}, the
electrons within the rings are moving along the lines of $B({\bf
  r})=0$ (black lines on the figure), so that the current density is
nonzero along the circle. One can see at this picture that the current
density has an oscillating fine structure in the direction
perpendicular to the lines $B({\bf r})=0$. Similar oscillations have
been observed by Hofstetter {\it et al}\cite{Hofstetter1996} for
electrons moving in linear magnetic fields.
%

\begin{figure}
  \centering
   \includegraphics[width=60mm]{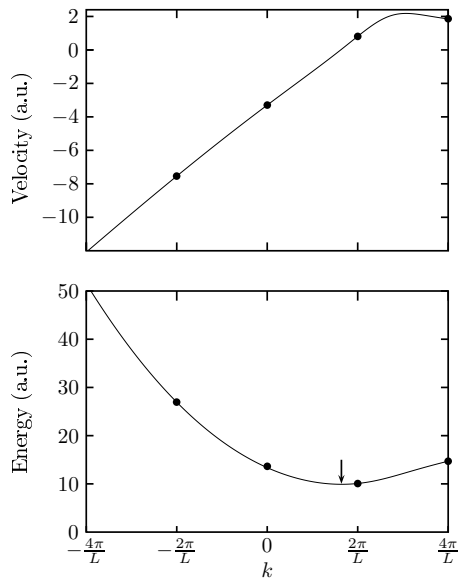}
   \caption{Schematic view of the velocity (a) and dispersion (b) curves
   (only the first band is shown) of an electron moving in a
     linear magnetic field. When the line $B({\bf r})=0$ is infinite,
     the energy spectrum is continuous as a function of momentum
     $k$ along the line. The minimum at the dispersion
     is indicated by arrow, and it corresponds to $k_{min}\ne 0$.
     For finite $L$ the energy spectrum is discrete (black points).}
  \label{fig:10}
\end{figure}

The appearance of persistent currents is due to the chirality of
electron motion in the nonuniform magnetic field along the line of
$B({\bf r})=0$. One can understand it by using a semiclassical
picture of the 1D motion in the inhomogeneous
field\cite{Muller1992}. The effective potential for the motion of
an electron along the zero-field line is different for the
opposite directions of the motion. It results in a strong
asymmetry of the electron energy spectrum with respect to $k\to -
k$, where $k$ is the electron momentum along the zero-field line.

In the case of 2D periodic field, this chirality of electron
energy spectrum should be combined with the fact that the
trajectories are closed in circles. Then the energy spectrum is
not only asymmetric but also quantized due to the quantization of
the motion along the circular trajectory. Using the semiclassical
picture, one can find the quantized values of the momentum from
relation $k_n -A_l=2\pi n/L$, where $A_l$ is the vector potential
along the contour and $L$ is its length. The circulation of $A_l$
along the circle is equal to the encompassed flux, which causes
the difference in phases for electron motion in opposite
directions. Thus, the condition of quantization can be also
presented as $k_n=2\pi(n+\phi /\phi _0)/L$.
In our 2D model of periodic field, the ratio $\phi /\phi _0$ can
be related to the parameter $\alpha $. The calculation of flux
through an isoline $B = 0$ using Eq.~(\ref{eq:10}) gives $\phi
/\phi _0\simeq 3.2437\alpha $.


Figure \ref{fig:10}-b shows schematically the energy spectrum as a
function of momentum $k$ along the circle. The points correspond
to the quantized values of $k$. As we see from Fig.~\ref{fig:10},
even for integer values of $\phi / \phi_0$, i.e., in absence of
the Aharonov-Bohm effect, the chirality of the motion in opposite
directions would result in appearance of nonzero electric current
along the circle.
We therefore have identified a novel mechanism for appearance of
persistent currents.

\section{Hall effect in the periodic magnetic field}

Here we consider the occurrence of non-vanishing off-diagonal
conductivity in the 2DEG with periodic magnetic field. As 
emphasized earlier, we assume that the average
magnetic field is zero, so that the ordinary Hall effect is absent. In
our case, the mechanism of nonzero Hall conductivity has the same
origin as the {\it "intrinsic
  mechanism"}\cite{Karplus1954,Jungwirth2002,M_Onoda2002} of the
anomalous Hall effect in ferromagnets, i.e., the Hall effect in 2DEG
in periodic magnetic field is related to the nontrivial {\it topology
  of electron energy bands} in the momentum space. However, unlike the
anomalous Hall effect in ferromagnets, it does not require any uniform
magnetization.

It should be noted that this effect is also quite different from
the recently proposed {\it "topological Hall effect"} in textured
ferromagnets.\cite{Bruno2004} Even though it was proposed in
Ref.~[\onlinecite{Bruno2004}] to use the periodic magnetic field
and 2D semiconductor with magnetic impurities, the only role of
the magnetic field was to order the magnetic moments in
correspondence to the field periodicity, so that the topological
properties of the magnetization profile are responsible for the
topological Hall effect. But in the model under consideration
there is no magnetization related to magnetic impurities.
Nevertheless, as we can see from the calculation of the
off-diagonal conductivity, $\sigma _{xy}$, the Hall effect is
nonzero.

The occurence of Hall effect and the quantization of Hall conductivity
has been discovered in the past in frame of 2D tight-binding honeycomb
model with an additional periodic magnetic field.\cite{Haldane1988}
The phase diagram of this model has two phases with Chern numbers
$\pm 1$. Also, the quantum Hall effect without any external magnetic
field has been found by Volovik in the model of electrons in $^3$He film.
\cite{Volovik1988} In both cases the origin of the Hall effect is related
to topological properties of electron energy bands.

Assuming that the Fermi level is in the energy gap and
in the absence of
impurities one obtains from the Kubo formula\cite{Jungwirth2002,M_Onoda2002}
\begin{equation}
  \label{eq:18}
  \sigma_{xy}=\frac{e^2}{\hbar}\sum_{n}
  \int \frac{d^2{\bf k}}{(2\pi )^2}\;
  f(\varepsilon_{n,\bf k})\, \Omega_{n,\bf k},
\end{equation}
where
\begin{equation}
  \label{eq:19}
  \Omega_{n,\bf k}=\nabla_{\bf k}\times \bmatA _{n,\bf k}
\end{equation}
is the Berry curvature, $\bmatA _{n, \bf k}=-i \langle n, {\bf
k}|\nabla_{\bf k}|n, {\bf k}\rangle$ is the gauge connection, and
$f(\varepsilon )$ is the Fermi function. Expression (\ref{eq:18})
was first found by Karplus and Luttinger\cite{Karplus1954} in the
context of the anomalous Hall effect.

Thus, if the Fermi level is in the energy gap and the temperature is
zero, the sum in Eq.~(\ref{eq:18}) can be presented as a sum over
fully occupied energy bands
\begin{equation}
  \label{eq:20}
  \sigma_{xy}=\frac{e^2}{\hbar}{\sum_{n}}^\prime
  \text{Ch}_n,
\end{equation}
where we denoted by
\begin{equation}
  \label{eq:21}
  \text{Ch}_n=\int \frac{d^2{\bf k}}{(2\pi )^2}\; \Omega_{n,\bf k}
\end{equation}
the Chern number of the $n$-th energy band. The Chern numbers are
integer\cite{Thouless1982,Niu1985} because they are topological
invariants\cite{chern}. Correspondingly, if the Fermi level is in
the gap, the Hall conductivity (\ref{eq:20}) is quantized like in case
of the quantum Hall effect\cite{Thouless1982}. The value of
$\sigma_{xy}$ changes when the gap between two occupied bands is
closing 
and depends explicitly on the dispersion relation around the points
where the band degeneracies occur\cite{Hatsugai1993,Oshikawa1994}. So,
the calculation of the off-diagonal conductivity reduces to the
calculations of the Chern numbers when the Fermi level is located
between two separated bands.

We calculate numerically the Berry curvature (\ref{eq:19}) and the
Chern numbers (\ref{eq:21}). It should be noted that the
computation of the derivative is a hard task because the phase of
the Bloch state is ill defined and gauge dependent. Moreover, the
summation over the first Brillouin zone involves a large number of
${\bf k}$-points. The problem of derivatives can be overcome by
expressing the Berry connection in term of the matrix elements of
the velocity operators:
\begin{equation}
  \label{eq:22}
  \Omega_{n,\bf k}= i \sum_{m\neq n} \frac{v_x^{nm} v_y^{mn}-v_y^{nm}
    v_x^{mn}}{(\varepsilon_{n,\bf k}-\varepsilon_{m,\bf k})^2},
\end{equation}
where
\begin{equation}
  \label{eq:23}
{\bf v}^{nm} =\left< n,{\bf k}\left| \frac{\partial
  \mathcal{H}_{\bf k}}{\partial {\bf k}}\right| m,{\bf k}\right>
\end{equation}
are the offdiagonal elements of the velocity operator.
This formula is gauge invariant but two difficulties remain,
(i) we need to calculate the sum over all energy bands and (ii)
the computational efforts to calculate the elements of the
velocity operator can still be important. In practice, the
summation over unoccupied bands is usually truncated over a few
number of unoccupied bands but a large number of ${\bf k}$ point is
still needed to calculate the Chern numbers.

\begin{figure}[h]
  \centering
  \includegraphics[width=8cm]{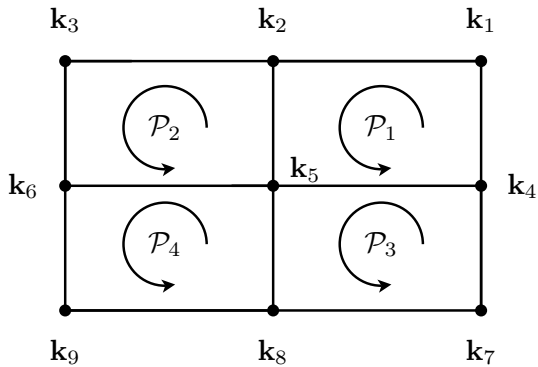}
  \caption{Principle of construction of the paths $\mathcal{P}_s$ in
    the case of a rectangular 
    Brillouin zone and a given mesh. The black
    points indicate the mesh taken for the decomposition of the first
    Brillouin zone. $\mathcal{P}_1={\bf k}_1 \to {\bf k}_2 \to {\bf
      k}_5 \to {\bf k}_4\to {\bf k}_1$ and the orientation is
    indicated by the arrow inside the rectangle. The others paths can
    be easy deduced using a similar construction.}
  \label{fig:11}
\end{figure}

Recently, Fukui {\it et al }\cite{Fukui2005} proposed another method
of calculation of the Chern numbers. This method has some
advantages: one needs to calculate only the Bloch states of occupied
bands over a coarse mesh of the first Brillouin zone. Moreover, the method
is gauge invariant. In order to calculate the Chern numbers, we
define the quantity
\begin{eqnarray}
  \label{eq:24}
  \gamma^n_{\mathcal{P}}=\text{Im} \log
  \left(\langle n,{\bf k}_1|n,{\bf k}_2\rangle
  \langle n,{\bf k}_2|n,{\bf k}_3\rangle
  \right. \nonumber\\ \times \left.
  \langle n,{\bf k}_3|n,{\bf k}_4\rangle
  \langle n,{\bf k}_4|n,{\bf k}_1\rangle\right) ,
\end{eqnarray}
where the function $\log z$ is defined in the complex plane
with branchcut along the negative real axis,
and $\mathcal{P}_s$ is a small closed path passing by the points
${\bf k}_s$ with $s=1,2,3,4$. The quantity
$\gamma^n_{\mathcal{P}}$, which is often called the field
strength, is the Berry phase that a Bloch state acquires when it
is transported adiabatically along the path $\mathcal{P}_s$. In
this formalism, the Chern number is given by a sum over the coarse
mesh of phases $\gamma^n_{\mathcal{P}_s}$:
\begin{equation}
  \label{eq:25}
  \text{Ch}_n =\sum_{\mathcal{P}_s}  \gamma^n_{\mathcal{P}_s}.
\end{equation}
The last step is to decompose the first Brillouin zone into small
paths $\mathcal{P}_s$ and to calculate the field strength
$\gamma^n_{\mathcal{P}_s}$ for each small path. The decomposition of
the first Brillouin zone is illustrated in Fig.~\ref{fig:11}.

\begin{figure}
\centering
\includegraphics[width=8cm]{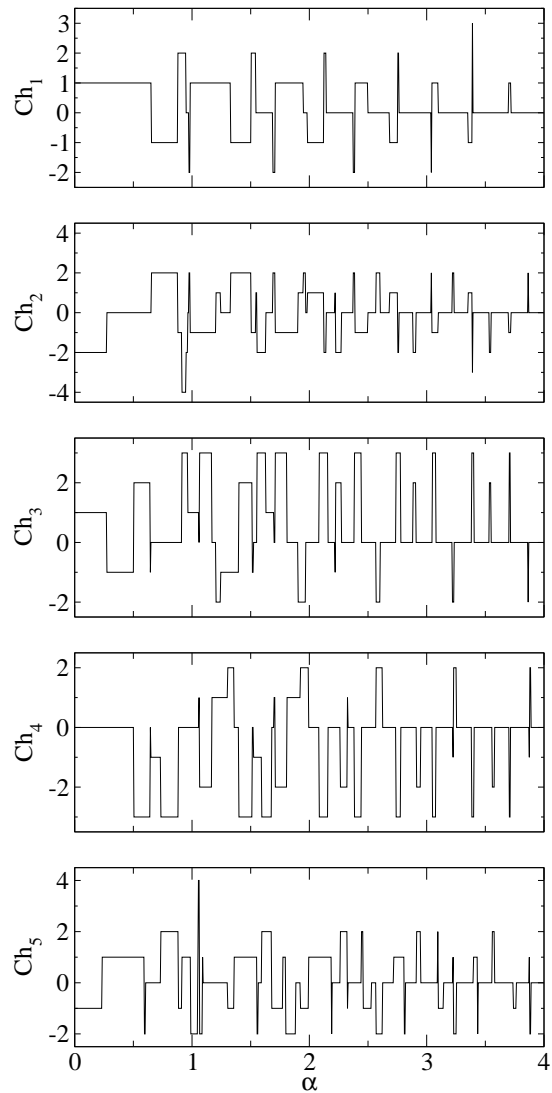}
\caption{Chern numbers of the first energy bands calculated as a
  function of the amplitude of the magnetic field $\alpha$. This
  figure shows that the Chern numbers have chaotic variations which
  are connected to closing gaps between the different bands.}
  \label{fig:12}
\end{figure}

We calculated the Chern numbers of the first five energy bands of our
problem using this method and obtained the results, which are shown in
Fig.~\ref{fig:12}. It should be noted that the sum given by
Eq.~(\ref{eq:25}) depends strongly on the number of planes waves used
to calculate the solutions of the Schr\"odinger equation. This
phenomena, which is not related to any numerical errors, occurs
because the truncation of the basis breaks the symmetry
$\mathcal{H}_{{\bf k}+{\bf g}}=\mathcal{H}_{\bf k}$, where ${\bf g}$
is a vector of the reciprocal space. The latter relation is used to
prove that the off-diagonal conductivity is an integer when the Fermi
level is in the gap. This effect can not be observed in the case of
tight-binding models because this symmetry is always verified. The
effect disappears quickly when the number of plane waves is
increasing.

The Chern numbers of the first five bands are presented in
Fig.~\ref{fig:12}. These results show that the variation of the
Chern numbers with the intensity of field looks rather chaotic and
there is no simple relation between the values of $\alpha$, for
which the band crossing occurs. It is known that the Chern numbers
are related to the band crossings between the different bands. The
value of the jump depends on the number of $k$-points in the first
Brillouin zone, where the band crossing occurs, and the dispersion
relation around these points\cite{Oshikawa1994}.
Although the physical mechanism is essentially the same, the
behavior observed here is much more complex as the one obtained in
Haldane's simplified model\cite{Haldane2004}.

According to Eq.~(\ref{eq:20}) the Hall conductivity can be found using
the Chern numbers, provided that the Fermi energy is within the gap.
As we can see from Fig.~\ref{fig:2}, the number of filled band changes
with $\alpha $. For example, if $\alpha =0.5$ then from
Fig.~\ref{fig:2}-b follows that two filled energy bands correspond to
the Fermi level located in the gap. Using the results of
Fig.~\ref{fig:12} we find ${\rm Ch}_1+{\rm Ch}_2=1$ and $\sigma
_{xy}=e^2/\hbar $.

Finally, we found that the Chern numbers tend to zero when $\alpha$ is
increasing. This is related to the low dispersive spectrum that we
obtained in Sec.~\ref{Sec:energy:spectrum}.

Our result shows that the off-diagonal
  conductivity is quantized if the Fermi level is located within the gap.
  It should be noted that this does not really prove the existence of Hall plateau,
  which could be observed when the field amplitude is varied.
  The plateau cannot be seen if the Fermi level jumps across the gap with changing
  field.
  A possible way to control smoothly the Fermi level position is to add impurities,
  which would lead to nonvanishing density of localized states in the gap.
  In this case the location of the Fermi level is stabilized in the gap, whereas
  the magnitude of plateau is determined by the interval of field corresponding to crossing
  the gap like in the case of quantum Hall effect in homogeneous field.

\section{Effect of electron spin}

In the previous consideration with did not take into account the spin of electron.
Now we discuss the effect of Zeeman coupling between
the electron spin and magnetic field. We show that in the case of strong magnetic field
it leads to a small correction of vector potential.

We consider the following Hamiltonian
\begin{equation}
  \label{eq:26}
  \mathcal{H}=\frac{1}{2m}
  \left( -i\hbar\nabla -q {\bf A}({\bf r})\right)^2
  + g\mu_B {\bf B}({\bf r})\cdot \bsig,
\end{equation}
where $g$ is the gyromagnetic constant, $\mu_B$ the Bohr magneton and
$\bsig $ are the Pauli matrices. It should be noted that $\mu _B$ is determined
by the mass of free electron, $m_0$, whereas the first term in (26) contains
the effective mass $m$. In the case of semiconductors like, e.g., GaAs,
the large difference of these masses, $m/m_0\ll 1$, leads to relatively
small Zeeman splitting.

After a local rotation $\mathcal{T}({\bf r})$ of the quantization axes along
the field in each point, the Hamiltonian reads
\begin{equation}
  \label{eq:27}
  \mathcal{H}=\frac{1}{2m}
  \left( -i\hbar\nabla -q {\bf A}({\bf r}) - q {\bf A}_g({\bf r})\right)^2
  + g\mu_B |{\bf B}({\bf r})|\sigma_z,
\end{equation}
with
\begin{eqnarray}
  \label{eq:28}
  {\bf A}_g({\bf r})=-\frac{i \hbar}{q} \mathcal{T}^\dagger({\bf r})\nabla
  \mathcal{T}({\bf r})={\bf A}^x_g \sigma_x+{\bf A}^y_g \sigma_y+{\bf A}^z_g\sigma_z.
\end{eqnarray}
The gauge field ${\bf A}_g({\bf r})$ depends only on unit vector ${\bf
  n}({\bf r})={\bf B}({\bf r})/|{\bf B}({\bf r})|$ which represents
the direction of field in point ${\bf r}$. In the adiabatic regime,
the off-diagonal terms of the gauge field ${\bf A}_g({\bf r})$ can be
neglected.\cite{Bruno2004} Hence, the spin up and down electrons are
decoupled, and the effective Hamiltonian describing each species have
a form similar to Eq.~(\ref{eq:1}) where the effective magnetic field
is given by ${\bf B}({\bf r})={\bf B}_z({\bf r})\pm{\bf B}_g({\bf r})$
with ${\bf B}_g({\bf r})=\nabla\times {\bf A}^z_g({\bf r})$.

The gauge field ${\bf B}_g({\bf r})$ depends only on geometry of the
magnetic lattice but does not depend on the amplitude of field ${\bf
  B}({\bf r}).$\cite{Bruno2004} Its flux is quantized and equal to
$2\pi n\phi_0$ with $n\in\mathbb{Z}$. Then the gauge field ${\bf
  B}_g({\bf r})$ can be neglected if the amplitude of magnetic field
${\bf B}({\bf r})$ is larger than $2\pi n\phi_0/S$ or, in others
terms, if $\alpha\gg 1$.


\section{Conclusions}

We calculated the energy spectrum of 2D electron gas in periodic field
with the symmetry of triangular lattice. As we can see, the
energy-band structure can be controlled by the variation of the
magnetic field strength. Using the realistic parameters of Fe
nanolattice with the lattice parameter of 100 nm, we have estimated
the magnitude of magnetic field as 5~kG.\cite{Bruno2004} This value
strongly depends on the gap between the nanolattice and the electron
gas, which gives a possible way to vary the field strength.

As a 2D electron gas one can use a metallic or semiconductor layer. In
this case there is also an additional crystal-lattice field but the
corresponding lattice constant $a_0$ is much smaller that the lattice
constant of the periodic magnetic field $a$. It means that we can
neglect the effect of periodic field of the crystal lattice as long as
the energy of electrons $\varepsilon \ll \hbar ^2/ma_0^2$.

We found that the low-energy electrons are effectively localized near
the lines of zero magnetic field, and in this state they produce an
equilibrium persistent currents in form of a ring array. The mechanism
of creation of such persistent currents is not necessarily induced by
the magnetic flux through the ring but is rather related to the
peculiarities of electron motion in opposite directions along the
zero-field lines.

We have also shown that the quantum Hall effect can be observed in
this system when the Fermi energy is located in the gap. In principle,
the problem of controlling the Fermi level location has been already
solved for 2D system with the gate. In a structure with fully
controlled periodic magnetic field and the Fermi energy, it would
result in a large functionality of the structure.

It should be noted that the impurity effects have been ignored in our
calculations. This is justified if the characteristic sizes of 2D
structure are smaller than the electron mean free path.

\section*{Acknowledgments}
This work is supported by the FCT Grant PTDC/FIS/70843/2006 in
Portugal, Polish Ministry of Science and Higher Education as
research projects in years 2006 -- 2009 and 2007 -- 2010, and by
the STCU Grant No.~3098 in Ukraine. VD thanks MPI f\"ur
Mikrostrukturphysik in Halle and the Institut N\'eel, CNRS/JFU, in
Grenoble for the hospitality.

\end{document}